\documentstyle[aps,12pt]{revtex}

\begin{document}
\title{Spin vs charge excitations in heavy-fermion compounds$^{\ast }$}
\author{R. J. Radwanski}
\address{Center for Solid State Physics, S$^{nt}$ Filip 5,31-150Krakow,Poland%
\\
Institute of Physics, Pedagogical University, 30-084Krakow, Poland}
\author{Z. Ropka}
\address{Center for Solid State Physics, S$^{nt}$ Filip 5,31-150Krakow,Poland%
\\
email: sfradwan@cyf-kr.edu.pl, http://www.css-physics.edu.pl}
\maketitle

\begin{abstract}
It is pointed out that the answer on the question about the role played by
spin and charge excitations will help to solve the physical origin of the
heavy-fermion phenomena. Our answer is that neutral spin-like excitations
are responsible for the heavy-fermion phenomena whereas the role of the
charge excitations is negligible.

PACS\ No: 75.20.H;

\end{abstract}

\date{(31.05.2002)}

\section{Introduction}

The microscopic origin of the heavy-fermion phenomena and the nature of
quasiparticles, despite of 25 years of very intensive theoretical and
experimental studies, is still a subject of controversy \cite%
{1,2,3,4,5,6,7,8,9}. The aim of this paper is to point out that neutral
spin-like excitations are responsible for the heavy-fermion phenomena
whereas the role of charge excitations is negligible (An extra remark - this
sentence has been underlined by the Chairman of SCES-02 in the rejected copy
with a note -WRONG!!).

\section{Theoretical understanding}

Characteristic heavy-fermion (h-f) phenomena like a large low-temperature
specific heat, a non-magnetic state (we would rather say a weakly-magnetic
state) anticipated from a Pauli-like low-temperature susceptibility and the
anomalous resistivity have been basically discussed in terms of the Fermi
liquid (FL). In the FL picture the excitations are charged and they become
possible due to the strong hybridization of \ $f$-electron and
conduction-electron states. In the FL picture the $f$ states lie at the
Fermi level and, as a consequence of the strong hybridization, the number of
the $f$ electrons, $n_{f}$, becomes not integer. The occupation number $%
n_{f} $ is the important factor in the FL description as many physical
properties are renormalized by the term (1-$n_{f}$)$^{-1}$. Kondo
temperature is directly related to this factor as one can read from Eq. 15
in Ref.\cite{3}, p.608. In this picture the disappearance of the local
moment is related to the unoccupied f states \cite{3}. Thus, in the FL
picture a deviation of the occupation number $n_{f}$ from the integer value
and the charge excitations play the fundamental role in formation of
heavy-fermion phenomena.

In our understanding of heavy-fermion phenomena we think that even in Ce, U,
Yb compounds, exhibiting the h-f phenomena, the $f$ states lie much below $%
E_{F}$ like in conventional rare-earth compounds \cite{10}. A large
low-temperature specific heat is related to the magnetic excitations well
understood in conventional rare-earth compounds in case of the Kramers
electron systems, i.e. systems with an odd number of the $f$ electrons. Such
the odd number electron system is realized in case of the $f^{1}$(Ce$^{3+}$%
), $f^{3}$ (U$^{3+}$), \ $f^{13}$ (Yb$^{3+}$) systems. All of them have the
Kramers doublet charge-formed (CF) ground state. This double degeneracy has
to be removed before the system reaches 0 K and, according to us, the h-f
compounds are compounds with extremely small magnetic temperatures. The
removal of the Kramers degeneracy is equivalent to the time-reversal
symmetry breaking (on the atomic scale) and to the formation of the magnetic
state (on the atomic scale). These phenomena can be well discussed within
the Quantum Atomistic Solid-State Theory (QUASST), that points out that $f$
atoms preserve much of their atomic properties becoming the full part of a
solid \cite{11}. In our model there is no single Sommerfeld coefficient $%
\gamma $ within the wide, but still below, say, 2 K, temperature range. In
contrary to a very loosely term ''non-Fermi liquid behaviour'' QUASST\
predicts that the ground state of the h-f compound is magnetic, i.e. with
the broken time-reversal symmetry. Moreover, it is the state with a low
local symmetry and the broken translational symmetry, in a sophisticated
manner, leading to the differentiation of the $f$ atoms with respect to the
shape of the CF\ ground state and of the magnetism. It causes that the
magnetic state is not coherent with respect to temperature, space and the
local moment direction, in the sense that local magnetic states, marked by
the splitting of single-ion Kramers doublet, appear at slightly different
temperatures at different sites and with different direction of the local
moment. According to this understanding the excitations are neutral,
spin-like and of very small energy. They mimics a spin-liquid with
non-trivial properties like the strong spin-lattice coupling and the
substantial orbital contribution to the magnetic moment. In the theoretical
description it will appear as the need for the attributing the spin with
unusual highly anisotropic properties. This fact causes that the local
excitations, with the reversal of the local moment, is not longer a local
event The Sommerfeld coefficient $\gamma $ can be extremely large at low
temperatures - Ropka has calculated $\gamma $ of 25 J/K mol \cite{12}. The
lower magnetic temperature, i.e. the temperature where the splitting of the
Kramers doublet appears, the larger $\gamma $ can be. It is very important
that our model overcomes the Nozieres exhaustion argument \cite{13} about
insufficient number of conduction electrons to compensate all localized
moments, of value comparable to that observed in the paramagnetic state, by
means of spin-compensation mechanism. We have shown that the crystal-field
(charge) mechanism, related to the anisotropic charge distribution at the
vicinity of the paramagnetic cation, is much more effective than the
spin-compensation mechanism \cite{14} in the reduction of the local magnetic
moments, even down to almost zero.Anisotropic charge distribution around the
4$f$-cation can produce the crystal-electric-field (CEF)\ ground state with
a quite small magnetic moment even in case of Kramers system.

\section{Discussion}

We take the growing evidence for the Non-Fermi -Liquid behaviour revealed in
rapidly growing number of compounds as the confirmation of our understanding
of the heavy-fermion phenomena. Our understanding, with the integer number
of $f$ electrons, concurs with theoretical results of Doradzinski and Spalek
who came to a number of $f$ electrons so close to 1 as 0.995 \cite{9}. It is
worth to remember that values for $n_{f}$ were given in years 1985-1990 as
about 0.70-0.80, see \cite{7}, for instance - thus a value of 0.995 we take
as practically 1. Within our model we have managed to describe
low-temperature specific heat of Nd$_{2-x}$Ce$_{x}$CuO$_{4}$ as originating
from excitations to the conjugate Kramers state of the Nd$^{3+}$ ions \cite%
{15} despite that Nd$_{2-x}$Ce$_{x}$CuO$_{4}$ has been announced in 1993 as
a new class of heavy-fermion superconducting compounds. We have managed to
describe an anomalous temperature dependence of the quadrupolar splitting in
YbCu$_{2}$Si$_{2}$ as the conventional crystal-field effect on the Yb$^{3+}$
ions \cite{16}. It was an important result as this anomalous dependence was
given as the conclusive evidence for the hybridization of $f$ electrons and
conduction electrons. Thus, the good description within the crystal-field
model has abolished the hybridization mechanism and that a n$_{f}$ value of
0.82 (hole) as completely artificial as Yb behaves in YbCu$_{2}$Si$_{2}$ as
the 4$f^{13}$ system. The existence of 3 $f$ electrons as highly-correlated 5%
$f^{3}$ system has been proved in heavy-fermion superconductor UPd$_{2}$Al$%
_{3}$ by observation of well defined CEF-like states \cite{17}.

QUASST can be applied not only to intermetallics, where the h--f behaviour
was found originally, but also to insulating rare-earth systems, known as
low-carrier systems (Sm$_{3}$Se$_{4}$, Yb$_{3}$S$_{4}$) and to nuclear
systems ($^{3}$He). Our model, developed already in 1994, has predicted the
possibility of h-f phenomena in 3$d$ and 4$d$ compounds - the h-f behaviour
has been discovered in LiV$_{2}$O$_{4}$ in 1997, indeed. QUASST predicts
smooth crossover from the heavy-fermion state to the conventional
localized-moment state with the Curie-Weiss law fulfilled, in agreement with
observations. The $f$ electrons, being localized in a number $n$ depending
on the partners and on the composition of a considered compound, are taking
the active part in the solid-state bonding, via the conventional coulombic
interactions.

Finally, we are at the Conference devoted to Strongly-Correlated Electron
Systems. In our atomic-like approach the correlations among the $f$
electrons are taken to be really very strong (thanks them, atomic-like terms
and multiplets as well as three Hund's rules are fulfilled).

\section{Conclusions}

We claim that heavy-fermion phenomena are caused by neutral spin-like
excitations, whereas the role of the charge excitations is negligible. The
ground state of heavy-fermion compounds is magnetic with the time-reversal
symmetry broken at the atomic scale.\bigskip\ 

An extra note added. We take a recent paper of Zwicknagl, Yaresko and Fulde,
Phys. Rev. B 65 (2002) 081103, that came to our attention during the SCES-02
Conference, with a novel model for heavy-fermion phenomena with two
localized $f$ electrons as confirmation of our 10-year understanding of
heavy-fermion phenomena with the importance of crystal-field states. In
particular, in the situation of Prof. P. Fulde who by twenty years advocated
for the itinerant $f$-electron origin of heavy-fermion phenomena. It is
worth to remind that in the SCES-94 Conference the crystal-field theory has
been rejected from the conference presentation as not at all related to
heavy-fermion physics. All members of the International Advisory Committee
have been informed about this abnormal situation but noone of them react in
order to fulfill normal scientific rules within the magnetic community.
Moreover, our CEF\ approach to heavy-fermion phenomena has been continuously
rejected by Fulde, acting as the Editor, from publication in Zeitsch. Physik
B (e.g. No MS606/94). Thus, we welcome this recent Zwicknagl et al.'s paper
admitting the existence of CEF states in heavy-fermion and uranium
compounds, though we think that the treatment of CEF in this paper is very
oversimplified. There still is a problem of the existence of the 5$f^{2}$ or
5$f^{3}$ system in UPd$_{2}$Al$_{3}$. In contrary to the 5$f^{2}$ system
considering by Zwicknagl et al. there is large evidence for the 5$f^{3}$
system. This evidence was published already in 1992, see references in Ref. %
\cite{17}, but Zwicknagl et al. have ignored it.        \ 

\ * This paper has been submitted 31.05.2002 to Strongly Correlated Electron
Conference in Krakow, SCES-02 getting a code NFL023. It has been presented
at the Conference, but has been rejected by the Chairman of SCES-02.

The paper has been given under the law and scientific protection of the
Rector of the Jagellonian University in Krakow, of \ the University of
Mining and Metallurgy and of Polish Academy of Sciences.

\end{document}